\documentclass[useAMS,usenatbib]{mn2e}

\usepackage{graphicx}
      
\def\Msun{M$_\odot$}
\def\msun{{\rm M}_\odot}

\def\lsim{\mathrel{\rlap{\lower 3pt \hbox{$\sim$}} \raise 2.0pt \hbox{$<$}}}
\def\gsim{\mathrel{\rlap{\lower 3pt \hbox{$\sim$}} \raise 2.0pt \hbox{$>$}}}

\title[Observational constraints on MBH
binary pairing]{Linking the fate of massive black hole binaries
to the active galactic nuclei luminosity function} 
\author[Dotti et al.]{M. Dotti$^{1,2}$\thanks{Massimo.Dotti@mib.infn.it}, A. Merloni$^{3}$, 
C. Montuori$^{4}$\\ 
$^{1}$ Universit\`a degli Studi di Milano-Bicocca, Piazza della
             Scienza 3, 20126 Milano, Italy\\
$^{2}$ INFN, Sezione di Milano-Bicocca, Piazza della
             Scienza 3, 20126 Milano, Italy \\
$^{3}$ Max Planck Institut f\"ur Extraterrestrische Physik, Giessenbachstrasse 1, D-85748 Garching bei M\"unchen, Germany\\   
 $^{4}$ Universit\`a degli Studi dell'Insubria, Via Valleggio 11,
             22100 Como, Italy\\            
 }
\begin{document}

\pagerange{\pageref{firstpage}--\pageref{lastpage}} 

\maketitle

\label{firstpage}

\begin{abstract}
Massive black hole binaries are naturally predicted in the context of
the hierarchical model of structure formation. The binaries that
manage to lose most of their angular momentum can coalesce to form a single
remnant. In the last stages of this process, the holes undergo an
extremely loud phase of gravitational wave emission, possibly
detectable by current and future probes. The theoretical effort
towards obtaining a coherent physical picture of the binary path down
to coalescence is still underway. In this paper, for the first time,
we take advantage of observational studies of active galactic nuclei
evolution to constrain the efficiency of gas-driven binary
decay. Under conservative assumptions we find that gas accretion
toward the nuclear black holes can efficiently lead binaries of any
mass forming at high redshift ($\gsim 2$) to coalescence within the
current time. The observed ``downsizing'' trend of the accreting black
hole luminosity function further implies that the gas inflow is
sufficient to drive light black holes down to coalescence, even if
they bind in binaries at lower redshifts, down to $z \approx 0.5$ for
binaries of $\sim 10^7 \msun$, and $z \approx 0.2$ for binaries of
$\sim 10^6 \msun$. This has strong implications for the detection
  rates of coalescing black hole binaries of future space-based
  gravitational wave experiments.
\end{abstract}


\section{Introduction}

Massive black hole (MBH) pairs are expected to form during galaxy mergers
\citep[][BBR, hereafter]{begelman80}. If the nuclei of the two
merging galaxies manage to survive against the tidal forces in play for long
enough \citep[e.g.][]{callegari09, vw14}, dynamical
friction can efficiently bring the two MBHs to the centre of the galactic
remnant, forcing them to bind in a MBH binary (BHB).

>From the binary formation on, dynamical friction becomes less and less
efficient (BBR), and other dynamical processes are needed to
further evolve the binary. In particular, the interaction with single stars
 and with nuclear gas have been thoroughly
investigated \citep[see][for a recent review]{dotti12}. The
sole effect of gravitational wave emission forces the two MBHs to coalesce
within the Hubble time if any physical process manages to shrink the
semi-major axes of the BHB down to:
\begin{equation}
\label{eq:agw} a_{\rm GW} \approx 2\times 10^{-3}
f(e)^{1/4}{q^{1/4}\over(1+q)^{1/2}}\left ({M \over   10^6\,\msun}\right )^{3/4}\,{\rm pc},
\end{equation}
where $M=M_1+M_2$ is the total mass of the binary, $q=M_2/M_1$ is
its mass ratio, and $f(e)=[1+(73/24)e^2+(37/96)e^4](1-e^2)^{-7/2}$ is a
function of the binary eccentricity $e$ \citep{peters64}. The assessment of how effective
the various processes are to evolve the binary down to $a_{\rm GW}$ is usually
referred to as the 'last parsec' problem.

Attempts to determine the fate of BHBs (whether they manage to reach $a_{\rm
  GW}$ and to coalesce or they remain bound in double systems forever) have
been made first considering gas-poor environments, where BHB
dynamics is assumed to be driven by three-body interactions with single stars.
In principle, only stars whose orbits intersect the BHB can efficiently
interact with it. In an extended stellar system, however, only a small
fraction of the phase space (the so called binary ``loss cone'') is populated
by such orbits.  Stars interacting with the BHB remove energy and angular
momentum from the binary, getting ejected from the loss cone. The binary evolution 
timescale is hence related to the rate at which new stars are
fed into the loss cone \citep[e.g.][]{mf04}. Physical mechanisms able
to efficiently refuel the loss cone are required in order for the binary to
coalesce in less that an Hubble time.  Possible mechanisms that have been
proposed so far are: the presence of massive perturbers \citep[such as giant
molecular clouds][]{mp08}, deviations from central symmetry
(e.g. Khan, Just \& Merritt 2011; Preto et al. 2011; Gualandris \& Merritt
2011, but see also Vasiliev, Antonini \& Merritt 2014) and gravitational
potential evolving with time \citep[e.g.][]{vam14}.

Similarly, the effect of the interaction between BHBs and nuclear gas
 has been explored analytically as well as numerically
\citep[see][for an up to date review]{dotti12}. While the full details
of the gas/binary interaction are still under debate, mainly due to
the complexity of the system, a clear issue remains to be
addressed. Similarly to the stellar-driven case, the migration
timescale of the BHB primarily depends on how much gas is able to
spiral toward the MBHs and interact with them, instead of, e.g. turn
into stars \citep[e.g.][]{lodato09}. This problem is remarkably
similar to the fueling problem of Active Galactic Nuclei (AGN),
i.e. how gas manages to lose most of its angular momentum in order to
sustain the observed nuclear activity.

Differently from the stellar driven case, however, observational studies of the
AGN population allow to constrain the properties of the gas flowing onto a MBH
(in particular its mass accretion rate).  Decades of
multi-wavelength surveys of accreting MBHs have provided a relatively
robust picture of the AGN luminosity function evolution \citep[see
  e.g.][]{Hasinger2005,Hopkins2007,Buchner2014}. Coupling such
evolution with the observationally determined MBH mass function via a
continuity equation \citep{Cavaliere71,Small92} allows to further
infer the evolution of the nuclear inflow rates as a function of MBH
mass and redshift (e.g. Merloni \& Heinz 2008, Shankar et
al. 2013). 

In this work we will assume that the fueling of BHBs is
consistent with that of single MBHs in the same mass and redshift
interval. In this way we can estimate the incremental reservoir of
angular momentum that a BHB can interact with during its cosmic
evolution, constraining the binary fate, at least in a statistical
fashion, directly from observations.

\section{Model}

\subsection{Gas driven BHB dynamics}
To get an order of magnitude estimate of the binary coalescence
timescale we propose a very simple zeroth-order model for the
interaction between a BHB and a circum-binary accretion disc. We
assume that the BHB is surrounded by an axisymmetric, geometrically
thin accretion disc co-rotating with the BHB. Under this assumption,
the gas inflow is expected to be halted by the binary, whose
gravitational torque acts as a dam, at a separation $r_{\rm gap}
\approx 2 a$ \citep[e.g.][]{al94}, where $a$ is the binary semi-major
axis. At this radius the gravitational torque between the binary and
the disc is perfectly balanced by the torques that determine the large
scale ($r \gg a$) radial gas inflow. Then, we can
write the variation of the binary orbital angular momentum magnitude
($\rm{L_{BHB}}$) as:
\begin{equation}\label{eqn:delta_mom1}   
{\rm dL_{\rm BHB} = -dL_{\rm gas}} = -\dot m \, {\rm d}t \, \sqrt{G \, M \,
  r_{\rm gap}}   
\end{equation}
where $M$ is the total binary mass, and $\dot m$ is the accretion rate
within the disc.  Considering the definition of $\rm{L_{BHB}}=\mu \,
  \sqrt{G\,M\,a}$, where $\mu$ is the binary reduced mass, from
eq.\,\ref{eqn:delta_mom1} we derive:
\begin{equation}\label{eqn:delta_a1}   
\frac{\mu}{2\sqrt{2}} \, \frac{{\rm d}a}{a} \approx - \dot{m} \, {\rm d}t,   
\end{equation}
where $\mu$ does not evolve in time consistently with the assumption that the
binary interacts with the disc only through the gravitational torques, that
stop the gas inflow preventing the binary components to accrete.  

As a note of caution, we stress that the circum-binary discs could be,
in principle, counter-rotating with respect to the BHBs they orbit
\citep{nixon11b}. In this configuration the gas interacts with the BHB
at $\approx a$, instead of $\sim 2a$ \citep{nixon11a}, and the
specific angular momentum transfer per unit time is uncertain,
depending on how strongly the secondary MBH is able to perturb the gas
inflow. As an example, if all the gas passing through $a$ bounds to
the secondary MBH, the binary angular momentum diminishes by two times
the angular momentum carried by the gas. In this scenario, the BHB
evolves on the same timescale regardless of the BHB-disc relative
orientation. Moreover, \cite{Constanze2014} demonstrated that BHBs
  embedded in self-gravitating retrograde discs may secularly tilt their
  orbital plane toward a coplanar prograde equilibrium configuration.
For these reasons we will focus our investigation on the
prograde case in the following.

As a first order of magnitude estimate of the coalescence timescale we
can make the simplifying (although quite common) assumption of an
Eddington limited accretion event and integrate
eq.\,\ref{eqn:delta_a1}.  Further assuming a fixed radiative
efficiency ($\epsilon =0.075$, see section 2.2) we obtain
\begin{equation}\label{eqn:delta_t1}   
  \Delta t_{\rm BHB} \sim \ln{\left(\frac{a_i}{a_c} \right ) \frac{\mu \,
      \epsilon \,c^{2}}{2\sqrt{2} \,L_{\rm Edd}}} \sim  10^{7} \,
  \frac{q}{(1+q)^2}\ln{\left(\frac{a_i}{a_c} \right ) \, {\rm yr} }     
\end{equation}
where $a_i$ and $a_c$ define the binary separation range where the MBH-gas
interaction drives the binary orbital decay.

Eq.~\ref{eqn:delta_t1} shows that, in order to
coalesce, the binary has to interact with an amount of matter of the order of
its reduced mass, with only a weak dependence on the exact ratio between the
initial and final separation. A conservative estimate for this ratio can be
obtained setting $a_i$ equal to the radius at which the two MBHs bind in a
binary
\begin{equation}\label{eqn:binary} 
a_i \sim G
  M/2\sigma_{\star}^2 \sim 0.5 ~\left(\frac{M}{10^6 \msun}\right)^{1/2} \, {\rm pc},
\end{equation}
where we have assumed the $M -\sigma_{\star}$ relation
\citep{gultekin}. The final separation can be conservatively estimated
as $a_c \sim 6\times 10^{-5} (M/10^6 \msun)^{3/4}$ pc, in order for
the BHB to coalesce due to the emission of gravitational waves in
$\sim 10^4 \,$ yr.  Under these assumptions $a_i/a_c \propto M^{-1/4}$
and $\ln(a_i/a_c) < 9$ for a binary mass $> 10^6 M_{\odot}$.

\subsection{Gas inflow onto BHBs: observational constraints}

We can now, for the first time, try to relax any a priori assumption on the
accretion rate in the circum-binary disc (such as the Eddington limit used in
eq.~\ref{eqn:delta_t1}), assuming an observational driven prescription
for the evolution of $\dot m$ as a function of MBH and redshift. In
particular, we adopt the average accretion rates obtained assuming
that the MBH evolution is governed by a continuity equation, where the
MBH mass function at any given time can be used to predict that at any
other time, provided the distribution of accretion rates as a function
of black hole mass is known. The continuity equation can be written as:
\begin{equation}
\label{eq:continuity}
\frac{\partial \psi(m,t)}{\partial t} +
\frac{\partial}{\partial m}\left( \psi(m,t) \langle \dot M
  (m,t)\rangle \right)=0
\end{equation}
where $m=Log\, M$ ($M$ is the black hole mass in solar units),
$\psi(m,t)$ is the MBH mass function at time $t$, and $\langle \dot
M (m,t) \rangle$ is the average accretion rate of a MBH of mass $M$
at time $t$. The average accretion rate can be defined through a
``fueling'' function, $F(\dot m,m,t)$, describing the distribution
of accretion rates for objects of mass $M$ at time $t$: $\langle \dot
M(M,z)\rangle = \int \dot M F(\dot m,m,z)\, \mathrm{d}\dot m$.
Such a fueling function is not known a priori, and observational
determinations thereof have been able so far to probe robustly only
the extremes of the overall population. However, the fueling
function can be derived by inverting the integral equation that
relates the luminosity function ($\phi$) of the AGN population with its
mass function. Indeed we can write:
\begin{equation}
\label{eq:filter}
\phi(\ell,t)=\int F(\ell-\zeta,m,t) \psi(m,t)\; \mathrm{d}m
\end{equation}
where we have called $\ell=Log\, L_{\rm bol}$ and $\zeta=Log\,
(\epsilon c^2)$, with $\epsilon$ the radiative efficiency. This is
assumed to be constant and its average value can be estimated by means
of the Soltan argument \citep{Soltan1982}, which relates the mass
density of remnants MBH in the local Universe with the integrated
amount of accreted gas during the AGN phases, as identified by the
luminosity function. 

\begin{figure*}
\includegraphics[scale=0.4]{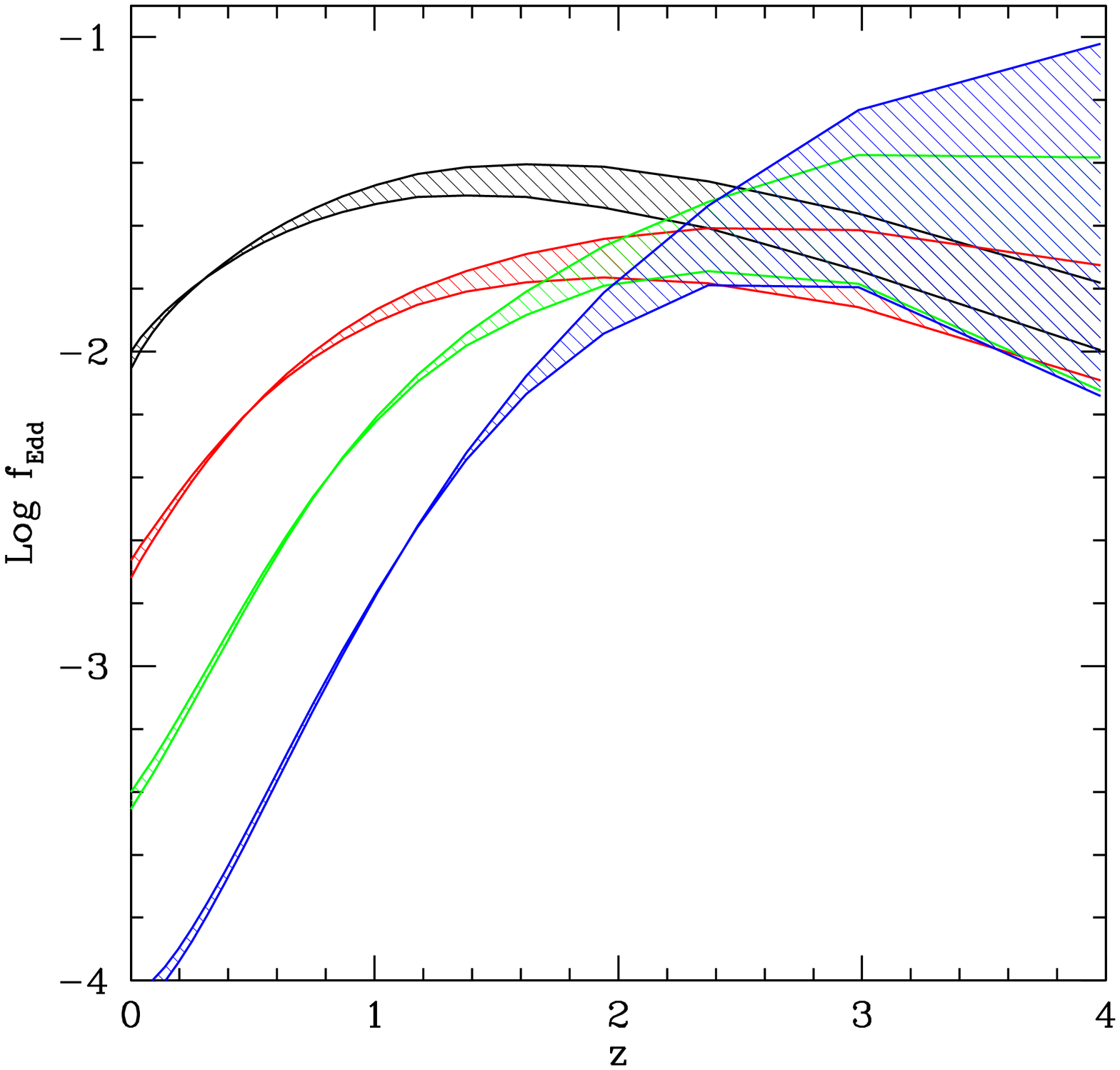}
\includegraphics[scale=0.4]{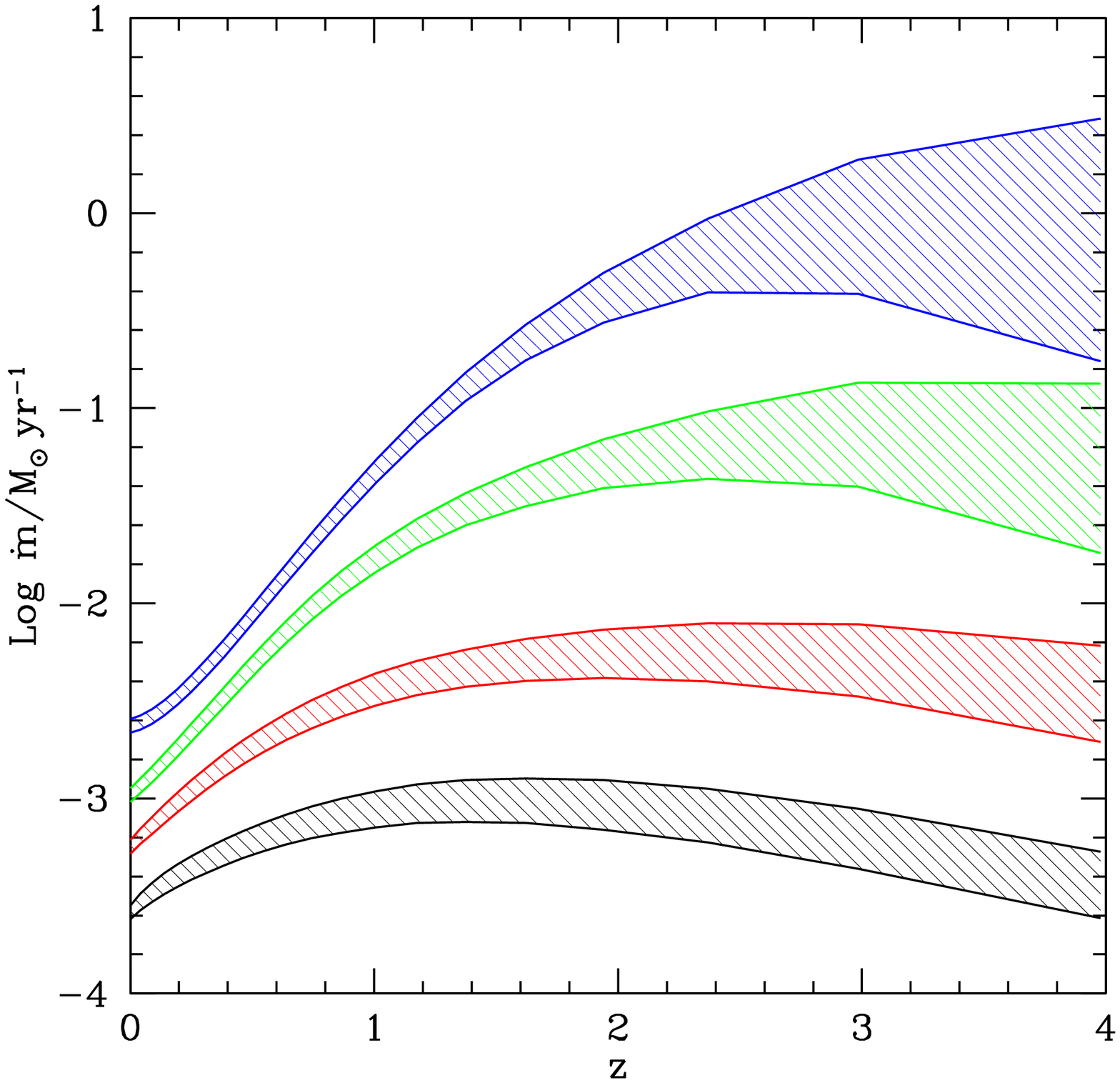}
\caption{Average Eddington ratios (left panel) and mass accretion
  rates (right panel) of MBHs as function of $z$. Black, red, green and blue colors
  refer to MBH masses of $10^6$, $10^7$, $10^8$, and $10^9$ \Msun,
  respectively. The shaded areas show the range of values comprised
  between the two limiting cases considered for the radiative
  efficiency (see discussion in the text) corresponding to
  $\epsilon = 0.075$ and $\epsilon = 0.1$.
}
\label{fig:mdot}
\end{figure*}

Gilfanov \& Merloni (2014) reviewed the most
recent assessments of the Soltan argument. Adopting as a starting
point the bolometric AGN luminosity function of Hopkins et al. (2007),
the estimate of the (mass-weighted) average radiative efficiency,
$\langle \epsilon \rangle$, can be expressed as:

\begin{equation}
\label{eq:soltan}
\frac{\langle \epsilon \rangle}{1-\langle \epsilon \rangle} \approx 0.075 \left[\xi_0(1-\xi_i-\xi_{\rm CT}+\xi_{\rm lost}) 
\right]^{-1} 
\end{equation}

where $\xi_0=\rho_{\rm BH,z=0}/ 4.2\times 10^5 M_{\odot} {\rm Mpc}^{-3}$ is
the local ($z=0$) MBH mass density in units of 4.2$\times 10^5 M_{\odot} {\rm
  Mpc}^{-3}$ \citep{marconi2004}; $\xi_i$ is the mass density of black holes
at the highest redshift probed by the bolometric luminosity function, $z
\approx 6$, in units of the local one, and encapsulates our uncertainty on the
process of BH formation and seeding in proto-galactic nuclei; $\xi_{\rm CT}$
is the fraction of the MBH mass density (relative to the local one) grown in
heavily obscured, Compton Thick AGN; finally, $\xi_{\rm lost}$ is the fraction
of BH mass contained in ``wandering'' objects, that have been ejected from a
galaxy nucleus, for example, in the aftermath of a merging event because of
the anisotropy in the emission of gravitational waves \citep[e.g.][and
  references therein]{Lousto13}. More recent estimates of the fraction of MBH
mass density accumulated in heavily obscured, Compton-Thick AGN
\citep{Buchner2014} suggest that $\xi_{\rm CT} \approx 0.35$. Neglecting
$\xi_i$ in eq.~\ref{eq:soltan} (i.e. assuming a negligibly small seed BH mass
density), the average radiative efficiency will vary approximately between
0.075 and 0.1 for $0< \xi_{\rm lost} < 0.3$. Therefore in the following we
will use the results obtained by performing a numerical inversion of
eq.~\ref{eq:filter}, based on a minimization scheme that used the Hopkins et
al. (2007) AGN bolometric luminosity function as a constraint, and assuming a
fixed radiative efficiency in the range $0.075<\epsilon<0.1$.  The average
Eddington ratios $f_{\rm Edd}$ (bolometric luminosity normalized to the
Eddington limit) and accretion rates obtained in this way are shown as a
function of redshift in the left and right panels of figure~\ref{fig:mdot},
respectively. We note that increasing the radiative efficiency value implies a
decrease in the average accretion rates, especially for higher MBH masses at
higher redshifts where the MBH evolution is relatively more important. This is
consistent with the adopted calculation scheme where the AGN luminosity
function is assumed as a constraint to derive the accretion rates estimates.

\section{Results}

\begin{figure}
\includegraphics[scale=0.4]{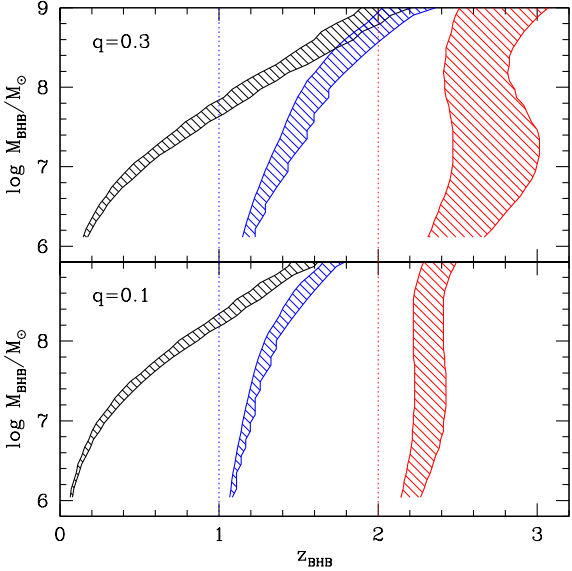}
\caption{Minimum redshift $z_{\rm BHB}$ at which two MBHs must bind in a
  binary of total mass $M_{\rm BHB}$ in order to coalesce within three
  different redshifts $z_{\rm coal}=0$ (black lines), 1 (blue lines)
  and 2 (red lines). The upper and lower panels correspond to a mass
  ratio $q=0.3$ and $q=0.1$, respectively. As in figure \ref{fig:mdot}, the
  shaded areas correspond to the range of values obtained considering
  a fixed radiative efficiency comprised between $\epsilon=0.075$
  (left curve) and $\epsilon=0.1$ (right curve). Thin dotted lines
  mark the three values assumed for $z_{\rm coal}$, and are shown to
  facilitate the reader. The case represented in the figure is when
  all the gas inflowing toward the galaxy nucleus interacts with the
  secondary MBH.}
\label{fig:zq}
\end{figure}

The observational constraints on the average value of $\dot{m}$ (function of M
and z) discussed in section 2.2 allow us to numerically integrate
eq.~\ref{eqn:delta_a1} to determine the migration timescale $\Delta t_{\rm
  BHB}$ for any BHB. We can further translate $\Delta t_{\rm BHB}$ into an
estimate of the minimum redshift $z_{\rm BHB}$ at which a BHB of mass $M$ must
form in order to coalesce within a given redshift $z_{\rm coal}$. The results
of the numerical integration are shown in figure~\ref{fig:zq} for a binary
mass ratio $q=0.1$ (lower panel) and $q=0.3$ (upper panel).
No 'last parsec' problem seems to exist for binaries of total mass $M<10^7$
\Msun and $q \lsim 0.3$ formed at $z_{\rm BHB} > 0.5$. More massive binaries
($M>10^8$) \Msun\, do not coalesce within the present time if formed at
$z_{\rm BHB} < 1.2$, and the extreme cases of $M>10^9$ \Msun\, coalesce
within $z=0$ only if formed at $z_{\rm BHB} \gsim 2$. Binaries forming at
higher redshift coalesce in shorter times, since the average accretion rates
increase with $z$ within the redshift interval considered in this
analysis. For example, all the $q<0.3$ binaries forming at $z_{\rm BHB} \sim 2.5$
coalesce within $z_{\rm coal}=2$.

\begin{figure}
\includegraphics[scale=0.4]{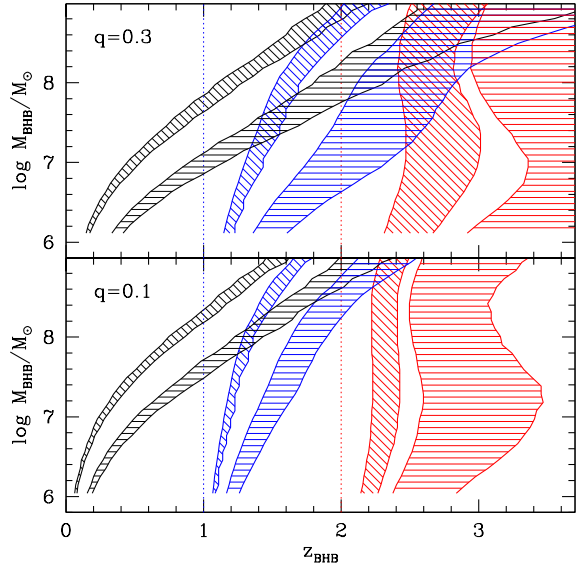}
\caption{Same as figure~\ref{fig:zq} either assuming that all the gas
  interacts with the secondary MBH (shading with inclined lines) or
  that only a fraction of 0.4 of the inflowing gas mass interacts with
  $M_2$ (shading with horizontal lines).}
\label{fig:z0.1}
\end{figure}

As a note of caution, we stress that the assumption that all the gas
inflow is stopped by the binary is oversimplifying. As a matter of
fact, numerical 2-D and 3-D simulations (independently of the exact
treatment of gravity or hydrodynamics) demonstrated that the
deviations from axisymmetry close to the binary, driven by its
gravitational potential, allow for periodic inflows of gas within
$r_{\rm gap}$ \citep{Hayasaki08, Cuadra09, Roedig11, Sesana12, Shi12, Noble12,
Dorazio13, Farris14}. To put firm upper limits to the
MBH migration timescale, we assume that only a fraction $f=0.4$ of the
gas inflow interacts dynamically with the binary, while the remaining
$60\%$ of the gas fails to strongly interact gravitationally with the
binary, and falls onto one of the MBHs unimpeded. A simple timescale
estimate for the binary shrinkage, under the assumption of MBHs
accreting at a fixed fraction of the Eddington limit, can be obtained
replacing $L_{\rm Edd}$ with $f\,L_{\rm EDD}$ in
eq.~\ref{eqn:delta_t1}. Reducing the fraction of interacting matter
increases the migration timescale, hence increasing the minimum
$z_{\rm BHB}$ required for the BHB to coalesce within a given $z_{\rm
  coal}$ as shown in figure~\ref{fig:z0.1}. 

As expected, decreasing $f$ the evolution of every binary slows down, but the
general trends discussed while commenting the $f=1$ cases remain valid. BHBs
with total mass $M\sim 10^7$ \Msun at $z > 2$ are
particularly affected, because of the redshift at which their typical
Eddington ratio peaks (see figure~\ref{fig:mdot}). Still, these BHBs manage to
coalesce between $z_{\rm coal}\approx 1 - 2$, as well as their more massive
counterparts.

\section{Conclusions}

We estimated the gas driven orbital decay of BHBs from the instant at which
they bind in a binary down to their final coalescence. For the first time we
propose an observationally driven approach, that has the advantage of not
being affected neither by any assumption on the (largely unknown) feeding process
driving the accretion, nor by the fraction of the gas inflow that turns into
stars at large scales before interacting with the BHB.

Our investigation proves that 1) high redshift BHBs of any mass coalesce on a
very short time; 2) Low mass BHBs ($M\lsim 10^7 \msun$) formed at low redshift
manage to merge anyway within $z=0$, since their accretion history peaks
at lower redshifts. These findings are particularly relevant since the
coalescence of low mass BHBs is one of the sources of gravitational waves
detectable by future space based gravitational wave interferometers, such as the
mission concept eLISA \citep{lisa}.

We have worked under very conservative assumptions: \\
- We assumed that binaries in the late stages of galaxy mergers
are fueled as much as MBHs in comparable isolated galaxies, without assuming
any merger driven boost in the accretion. The merger process itself is
considered, however, an efficient reshuffler of the gas angular
momentum at galactic scales, driving efficiently gas inflows all the
way down to the two MBHs, as confirmed by observations
\citep[e.g.][]{kk84,k85,a07,k11,e11,s11,s14} as well as by a wealth of
numerical works performed on different kind of mergers
\citep[e.g.][]{dsh05,jbn09,hopkins10,callegari11, vw12,capelo14}.\\
- We have assumed that all the gas accretion onto MBHs is
radiatively efficient, and that the mass accretion rate at few gravitational
radii ($R_g$, where basically all the luminosity is emitted) is equal to the
that at thousands of $R_g$, where the gas interacts with the secondary MBH. We
stress that a significant fraction of the accretion flow, however, could be
ejected in the form of fast outflows, as often found in numerical simulations
\citep[e.g.][]{proga03, narayan12}.

Under our conservative assumptions, gas driven migration of high mass ($M
\gsim 10^8 \msun$) BHBs formed at low redshift could be inefficient.  Such
binaries are of particular interest, being the only ones observable through
pulsar timing \citep[][and references therein]{Hobbs10}. The morphological and
dynamical characteristic of their hosts suggest, however, that interactions
with stars could play a significant role in the binaries shrinking. The hosts
of very massive MBHs often show triaxial profiles \citep[e.g.][and references
  therein]{faber97,kormendy09}. The lack of spherical and axial symmetry in
the potential of the hosts allows the single stars to modify substantially
their angular momentum components. Stars can hence re-fill the loss cone of
the binaries at rates significantly higher than those expected in spherical
systems\footnote{In spherical potentials the collisional refilling of the
  binary loss cone can lead binaries to coalescence within 
$10^{10}$ yr only
  if the total mass of the binary is $M \lsim 10^6 \msun$, see
  e.g. Section~8.3 in \cite{merritt13}.}, possibly leading BHBs to a fast
coalescence \citep[see][for a recent discussion]{vasiliev14}.

\section*{Acknowledgments}
We acknowledge the anonimous Referee, Alberto Sesana and Eugene Vasiliev for
useful comments and fruitful discussions.

\bsp

\label{lastpage}

\end{document}